\documentclass [12pt,a4paper,leqno]{article}
\title{The Integral Formulae for a Cylindrical Black Hole}
\author{M.D.IFTIME\thanks{MTIME@maths.qmw.ac.uk}}
\date{May 2002}
\begin{document}
\maketitle

\begin{abstract}
Here we derive the mass formulae for a cylindrical
black hole solution with positive cosmological constant $\Lambda >0$ and surrounded by dust. 
The expressions are generalising those found by Smarr for the mass and momentum of a Kerr black hole. 

\end{abstract}

 \vspace{2cm}  

\pagebreak

\section{Introduction}

\hspace{.5cm} Black hole are astrophysical objects that appear as a final state in a 
gravitational colapse.

Black holes are naturally studied as exact solutions of Einstein equations. Their theoretical proprieties, such us their stability, the no hair theorem, the physics of matter surrounding them, have been investigated in great depth and detail. However, almost all
these studies have concentrated on isolated black holes characterized which are stationary (static or axisymmetric) and asymptotic flat. Once black holes have became important to physics too, one should study them in realistic situation in which the black hole is
associated with a non-flat background.
However, we live in a universe with positive cosmological constant. Therefore, it would be interesting to consider the cases when the black hole is surrounded by matter and is embedded 
in a cosmological spacetime with $\Lambda >0$. 
In \cite{ls:cblk} is given a cylindrical black hole embedded in the anti de Sitter spacetime.
In \cite{mm} is investigated a static black hole embedded in Einstein universe. In \cite{CQG}is studied a cylindrical symmetric solution with positive cosmological and dust term that approaches Einstein universe on the axis of rotation.
Here we are concentrating on the conservation laws for a cylindrical black hole with 
$\Lambda >0$ and surrounded by dust.
In a black hole with a positive cosmological constant we can define another kind of event horizon.
A cosmological event horizon is defined as the boundary of the region from which light can never reach the external observer, due to the expansion effect caused by the repulsive $\Lambda $ term.

We generalize naturally the integral formulae found by Smarr for Kerr black hole to the case of a cylindrical black hole with $\Lambda >0$ and surrounded by dust.

For a stationary space-time, from the conservation law for the energy-momentum tensor 
$T^{ab}_{;b}=0$ we obtain the equation:

\begin{equation}\label{eq:1}
(k_{a}T^{ab})_{;b}=0=[(T^{ab}-1/2 g^{ab}T)k_{a}]_{;b}=(R^{ab}k_a)_{;b}
\end{equation}
where $k^a$ denotes the temporal Killing vector. Therefore the following quantity is conserved:
\begin{equation}\label{eqm1}
M_{k}=\int_{S_3}(R^{ab}k _a){\rm d}\sigma_a ={\rm const.}
\end{equation}
The following relation is true for any Killing vector $k$:

\begin{equation}\label{eqm2}
\left \{
\begin{array}{l}
k^{a;b}_{;b}=R^a_b k^b\\
k_{(a;b)}=0
\end{array}
\right.
\end{equation}
unde  $R_{ab}=R^c_{acb}$ is the Ricci tensor.

If we integrate the first relation on a spatial compact 3-surface $S$ with 
frontier $\partial S$ we obtain:
\begin{equation}\label{eq:1}
\oint_{\partial S} k^{a;b}{\rm d}f_{ab}=\int_S R^a_bk^b {\rm d}\sigma_a,
\end{equation}
where 
$\textrm{d}f_{ab}$ \c{s}i $\textrm{d}\sigma_a$ 
are the surface element on $\partial S$ and on $S$, respectively.
Now if we introduce (\ref{eqm1}) and (\ref{eq:1}) into Einstein's field equation
$R_{ab}-{1\over 2}Rg_{ab}+\Lambda g_{ab}=1/8\pi  T_{ab}$, 
we obtain the following expression:

\begin{equation}\label{masa}
M_{k}=-1/8\pi  \int_S (T^{a}_{b}-2T\delta ^{a}_{b}) k_{a}{\rm d} \sigma_b  +M_{\partial S}.        
 \end{equation}

\section{The Integral Formulae of a cylindrical black hole with $\Lambda >0$ surrounded by dust.}

We will use the above relations to derive the conservation laws for the mass, energy and 
flux of energy of a cylindrical balck hole with $\Lambda >0$ surrounded by dust.

For a stationary black hole with cylindrical symmetry there exist three Killing vector fields:
$\xi $, $\eta$ \c{s}i $\zeta$. Using \ref {eqm1} we obtain that the following quantities: mass $M$, momentum $J$ and energy flux $P$ are conserved,

\begin{equation}\label{Eq:1}
\begin{array}{l}
M=\int_{S_3}(R^{ab}\xi _a){\rm d}\sigma_a ={\rm const.}\\
J=\int_{S_3}(R^{ab}\eta  _a){\rm d}\sigma_a ={\rm const.}\\
P=\int_{S_3}(R^{ab} \zeta_a){\rm d}\sigma_a ={\rm const.}
\end{array}
\end{equation}
The above integrals are taken on a 3-surface $S_3$ with a frontier $\partial S_{3}\neq 0$, which intersect both black hole event horizons and it consists of the event horizon $\partial B$, the cosmological even horizon \cite{H} $C$, and the frontier at infinity $S_\infty $.

From(\ref{masa}) we can deduce a relation for the total mass $M$ of the gravitational system consisting of black hole and its surrounding dust:

\begin{equation}\label{Eq:2}
M_{total}=-1/4\pi \oint_{\partial S_\infty} \xi ^{a;b}{\rm d}f_{ab}=1/4\pi \int_{S_{\rm ext}} R^a_b \xi^b {\rm d}\sigma_a.
\end{equation}

We take $t $, the parameter on the integral curves of the Killing vector field
$\xi$, $\xi^a=\partial_{t}$. Then we define the Killing vector 
$\displaystyle l^{a}=\frac{\rm d x^a}{\rm d t}$, which is null on both the event horizon and cosmological horizon of the black hole.\footnote {The null Killing vector $l^a$ with respect to both horizons for the case of a static black hole in an asymptotically anti de Sitter spacetime coincides with $\xi^a$ \cite{HG}}
Because $l$ is vector Killing we can therefore write it as a liniar combination of 
$\xi $, $\eta$ \c{s}i $\zeta$,
$ l^a =\xi ^a+\Omega \eta^a +v \zeta ^a $. 
Here $\Omega $ is interpreted as the constant angular velocity and  $v$, the
translational velocity along axis of rotation $ \eta=0$ of the black hole.

Also we have ${\rm d} f^{1}_{ab} = l_{[a } n_{b]}{\rm d} A_{B}$, the surface element on the event horizon $\partial B$ and 
${\rm d} f^{1}_{ab} = l_{[a } m_{b]}{\rm d} A_{C}$, the surface element on the cosmological event horizon $\partial C$, where we denoted with 
$A_B$ the area of intersection of $S_{3}$ with the event horizon, $A_B$ the area of 
of cosmological event horizon on $S_{3}$ and $n^a$, respectively $m^a$, the null vectors orthogonal on event horizons.  

We therefore obtain the following formula for the integral mass of a cylindrical black hole with positive cosmological constant $\Lambda >0$:

\begin{equation}
\begin{array}{l}
 M_{total}-2\Omega _H J_H +\frac{\kappa_{B}} {4\pi}A_{B}+\frac{\kappa_{C}} {4\pi}A_{C}
 -2v_{H} P_H  \\
=\int_{S} (2T^a_b-T\delta ^a_b)\,\xi ^a{\rm d}\sigma _a. 
\end{array}.
\end{equation}
where we denoted by $k_B$ and $k_C$ the surface gravities for the black hole, which show to be 
constant on the black hole event horizon, respectively cosmological event horizon.

For a black hole asymptotic flat the total measured from infinity is $M_{total}=-1/4\int{\partial S_{\infty }}{\xi^{a;b}\mathrm{d}\sigma _{ab}}$.

The following quantities:

 \begin{equation}
\begin{array}{l}
M_H=-1/4\pi \oint_{\partial B}(\xi ^{a;b}){\rm d}f_{ab} =-1/4\pi \oint_{\partial B}(\xi ^{a;b})l_{[a } n_{b]}{\rm d} A\\
J_H=1/8\pi \oint_{\partial B}(\eta ^{a;b}){\rm d}f_{ab} =1/8\pi \oint_{\partial B}(\eta^{a;b})l_{[a } n_{b]}{\rm d} A\\

P_H=1/8\pi \oint_{\partial B}(\zeta ^{a;b}){\rm d}f_{ab}=1/8\pi \oint_{\partial B}(\zeta^{a;b})l_{[a } n_{b]}{\rm d} A
\end{array}
\end{equation}
represent the mass, final momentum and the flux of energy through the event horizon of black hole.

The relations (\ref{eq:1}) for the Killing vector $\zeta = \partial_z$ associated with translations along the axis of rotation take the form:

\begin {equation}\label{generalzation}
1/8\pi \int_S T^{ab} \zeta _a{\rm d}\sigma_b=\oint_{\partial B}\zeta ^{a;b}{\rm d}f_{ab}+
\oint_C \zeta ^{a;b}{\rm d}f_{ab}
\end {equation}

We can define $P=1/8\pi \int_S T^{ab} \zeta _a{\rm d}\sigma_b$ as the energy flux of the matter between the event and cosmological horizons cele dou\u{a} orizonturi calculated on the spacial hipersurface $S$. The second term, 
$P_C=\oint_C \zeta ^{a;b}{\rm d}f_{ab}$ defines the energy flux radiated through the cosmological horizon $C$.

The energy-momentum tensor is defined by $T_{ab}=\mu u^a u^b$, where $\mu $ is the density of the dust which rotates rigidly around the black hole with a 4-velocity
 $u^a=\xi ^a+\Omega \eta ^a+w\zeta^a$.

For the case of vacuum black hole with $T_{ab}=0 $, $\Lambda >0$ and axially symmetric we re-obtain the relations discovered by Smarr for a Kerr black \cite{christod}, \cite{smarr},

 \begin{equation}
\begin{array}{l}
\displaystyle A_{B}=8\pi [M^{2}+(M^{2}-J_{H}^{2})]^{1/2}\\
\displaystyle \Omega =\frac{4\pi J_{H}}{MA_{B}},\\ 

\displaystyle \frac{k}{4\pi }=\frac{(M^{4}-J_{H})^{1/2}}{2MA_{B}}
\end{array}
\end{equation}

\section{Conclusion}
In this paper we have deduced the expressions for the mass integral for a 
stationary cylindrical black hole with positive cosmological constant and dust term, expressions that generalize the mass formulae
found by Smarr for a Kerr black hole.

\end{document}